\documentclass[conference]{IEEEtran}

\usepackage{cite}
\usepackage{amsmath,amssymb,amsfonts}
\usepackage{algorithmic}
\usepackage{graphicx}
\usepackage{textcomp}
\usepackage{verbatim}
\usepackage{color}
\def\BibTeX{{\rm B\kern-.05em{\sc i\kern-.025em b}\kern-.08em
    T\kern-.1667em\lower.7ex\hbox{E}\kern-.125emX}}
    
\usepackage{booktabs} % For formal tables

\begin{document}

\title{A Historical and Statistical Study\\ of the Software Vulnerability Landscape
}

%\begin{comment}
\author{

\IEEEauthorblockN{Assane Gueye}
\IEEEauthorblockA{\textit{CMU-Africa} \\
Kigali, Rwanda \\
%a.delaitre@prometheuscomputing.com}
assaneg@andrew.cmu.edu}
\and

\IEEEauthorblockN{Peter Mell}
\IEEEauthorblockA{\textit{National Institute} \\
\textit{of Standards and Technology}\\
Gaithersburg MD, USA \\
peter.mell@nist.gov}

}
%\end{comment}

%\renewcommand{\shortauthors}{P. Mell et al.}

\maketitle

\begin{abstract}
Understanding the landscape of software vulnerabilities is key for developing effective security solutions. Fortunately, the evaluation of vulnerability databases that use a framework for communicating vulnerability attributes and their severity scores, such as the Common Vulnerability Scoring System (CVSS), can help shed light on the nature of publicly published vulnerabilities. In this paper, we characterize the software vulnerability landscape by performing a historical and statistical analysis of CVSS vulnerability metrics over the period of 2005 to 2019 through using data from the National Vulnerability Database. We conduct three studies analyzing the following: the distribution of CVSS scores (both empirical and theoretical), the distribution of CVSS metric values and how vulnerability characteristics change over time, and the relative rankings of the most frequent metric value over time. Our resulting analysis shows that the vulnerability threat landscape has been dominated by only a few vulnerability types and has changed little during the time period of the study. The overwhelming majority of vulnerabilities are exploitable over the network. The complexity to successfully exploit these vulnerabilities is dominantly low; very little authentication to the target victim is necessary for a successful attack. And most of the flaws require very limited interaction with users. However on the positive side, the damage of these vulnerabilities is mostly confined within the security scope of the impacted components. A discussion of lessons that could be learned from this analysis is presented.
\end{abstract}


\section{Introduction}

Understanding the landscape of software vulnerabilities is a key step for developing effective security solutions. It is difficult to counter a threat that is not well understood. Fortunately, there exist vulnerability databases that can be analyzed to help shed light on the nature of publicly published software vulnerabilities. The National Vulnerability Database (NVD) \cite{NVD} is one such repository. NVD catalogs publicly disclosed vulnerabilities and provides an analysis of their attributes and severity scores using the Common Vulnerability Scoring System (CVSS) \cite{CVSS}. CVSS is used extensively by security tools and databases and is maintained by the international Forum of Incident Response and Security Teams (FIRST) \cite{FIRST}.

The CVSS provides a framework for describing vulnerability attributes and then scoring them as to their projected severity. The attributes are metric values that are the input to a CVSS equation that generates the score. It is the vulnerability attribute descriptions (the metric values) that are of primary interest to our work, although we also look at the raw scores. The use of CVSS by vulnerability databases provides a suite of low level metrics, encapsulated in a vector, describing the characteristics of each vulnerability. CVSS was initially released in 2005 \cite{schiffman2004common}, was completely revamped with version 2 (v2) in 2007 \cite{mell2007complete}, and was updated with new and modified metrics in 2015 with the release of version 3 (v3) \cite{CVSSv3.0}\footnote{Minor update version 3.1 was released in 2019 \cite{CVSSv3.1} but the changes therein do not effect our work.}. The software flaw vulnerability landscape was thoroughly analyzed in the scientific literature using v2 when it was first released \cite{mell2007improving, schiffman2004common, mell2006common, holm2015expert, scarfone2009analysis, wang2011improved, holm2012empirical}, but little work has been done since to evaluate changes to that landscape over time. Also in our literature survey, we did not find a single study that uses the updated and significantly modified v3 to understand the software vulnerability landscape.

In this paper, we use the CVSS v2 and v3 data provided by the NVD to undertake a historical and statistical analysis of the software vulnerabilities landscape over the period 2005 to 2019. More precisely, we conduct three studies analyzing the following:
\begin{itemize}
    \item score distributions, 
    \item metric value distributions,
    \item and relative rankings of the most frequent metric values.
    %\item and the most prevalent patterns of co-occurrence of the metric values.
\end{itemize}

For our first study, we analyze and compare the distributions of CVSS v2 and v3 scores as generated from the NVD data. We then compare the empirical distributions against the theoretical score distributions, assuming that all CVSS vectors are equally likely (which is not the case, but it is illustrative to evaluate the differences).

For our second study, we compute the distributions of the CVSS metric values (i.e., vulnerability characteristics) for each year. We then analyze the differences from 2005 to 2019 to determine if and how vulnerability characteristics change over time. 

For our third study, we identify the most frequent metric values and analyze their relative rankings from 2015 to 2019. For each year and for both CVSS versions, we compute the values of the top 10 observed vulnerability metrics as well as their frequencies. We then generate parallel coordinates plots showing the values and frequencies of each metric for each year.

Our analysis shows that the software vulnerability landscape has been dominated by only a few vulnerability types and has changed very little from 2005 to 2019. 
For example, the overwhelming majority of vulnerabilities are exploitable over the network (i.e., remotely). The complexity to successfully exploit these vulnerabilities is dominantly low while attackers are generally not required to have any level of prior access to their targets (i.e., having successfully authenticated) in order to launch an attack. And most of the flaws require very limited interaction with users. On the positive side, the damage of these vulnerabilities is mostly confined within the security scope of the impacted components. Few vulnerabilities obtain greater privileges than is available to the exploited vulnerable component. 
%Lastly, the association rule mining analysis shows groups of vulnerability metrics that tend to always occur together. 

Our findings are consistent to previous studies \cite{mell2007improving} (mainly based on CVSS version 2). This indicates that the same vulnerabilities are still being found in our software, suggesting that the community has not been doing a great job correcting the most common vulnerabilities.  

The remainder of this paper is organized as follows. Section \ref{sec:CVSS-Dataset} presents the CVSS data sets that constitute the basis of our study. Section \ref{sec:Data-Anal} gives the details of our analysis and our discussion. Section \ref{sec:relatedWork} provides a summary of related work and Section \ref{sec:conclusion} concludes.

%********************************************************************

\section{The CVSS Datasets}
\label{sec:CVSS-Dataset}

CVSS consists of three metric groups: base, temporal, and environmental. The base group represents the intrinsic qualities of a vulnerability that are constant over time and across user environments, the temporal group reflects the characteristics of a vulnerability that change over time, and the environmental group represents the characteristics of a vulnerability that are unique to a user's environment \cite{CVSSv3.0}. In this work, we evaluate only the base metrics as no extensive database of temporal scores exists and the environment metrics are designed for an organization to customize base and temporal scores to their particular environment. 

Tables \ref{table:CVSS-v2} and \ref{table:CVSS-v3} show the base score metrics and possible values for v2 and v3, respectively. A particular assignment of metric values is then used as input to the CVSS base score equations to generate scores representing the inherent severity of a vulnerability in general apart from any particular environment. The raw score in the range from 0 to 10 is then often translated into a `qualitative severity rating scale' (None: 0.0, Low: 0.1 to 3.9, Medium: 4.0 to 6.9, High: 7.0 to 8.9, and Critical: 9.0 to 10.0) \cite{CVSSv3.0}.

Vulnerability analysts apply the metrics to vulnerabilities to generate CVSS vector strings. The vectors describe the metric values, but not the CVSS scores, for a particular vulnerability using a simplified notation.

\begin{table}[]
\centering
\caption{CVSS v2 metrics}
\begin{tabular}{|l|l|} \hline
%\multicolumn{2}{|c|}{CVSS V2}         \\  \hline
CVSS v2 Metrics & Metric Values                \\  \hline  \hline
Access Vector (AV)     & Network (N), Adjacent (A), Local (L) \\  \hline
Attack Complexity (AC) & Low (L), Medium (M), High (H) \\  \hline
Authentication (Au)    & Multiple (M), Single (S), None (N) \\ \hline
Confidentiality (C)    & Complete (C), Partial (P), None (N) \\ \hline
Integrity (I)          & Complete (C), Partial (P), None (N) \\ \hline
Availability (A)       & Complete (C), Partial (P), None (N)  \\ \hline
\end{tabular}
\label{table:CVSS-v2}
\end{table}

\begin{table}[]
\centering
\caption{CVSS v3 metrics}
\begin{tabular}{|l|l|} 
\hline
%\multicolumn{2}{|c|}{CVSS V3}  \\ \hline
CVSS v3 Metrics  & Metric Values      \\\hline\hline
Attack Vector (AV) & Network (N), Adjacent (A),
\\ & Local (L), Physical (P) \\\hline
Attack Complexity (AC)   & Low (L), High (H)  \\ \hline
Privileges Required (PR) & None (N), Low (L), High (H)   \\ \hline 
User Interaction (UI)    & None (N), Required (R)  \\ \hline
Scope (S)                & Unchanged (U), Changed (C)  \\ \hline 
Confidentiality (C)      & High (H), Low (L), None (N)\\ \hline
Integrity (I)            & High (H), Low (L), None (N) \\ \hline
Availability (A)         & High (H), Low (L), None (N) \\ \hline
\end{tabular}
\label{table:CVSS-v3}
\end{table}

\begin{comment}
\begin{table}[]
\centering
\caption{CVSS v2 metrics}
\begin{tabular}{|l|l|} \hline
\multicolumn{2}{|c|}{CVSS V2}         \\  \hline
Metric & Metric Values                \\  \hline  \hline
Access Vector (AV)     & Network (N), Adjacent (A), Local (L) \\  \hline
Attack Complexity (AC) & Low (L), Medium (M), High (H) \\  \hline
Authentication (Au)    & Multiple (M), Single (S), None (N) \\ \hline
Confidentiality (C)    & Complete (C), Partial (P), None (N) \\ \hline
Integrity (I)          & Complete (C), Partial (P), None (N) \\ \hline
Availability (A)       & Complete (C), Partial (P), None (N)  \\ \hline
\end{tabular}
\label{table:CVSS-v2}
\end{table}

\begin{table}[]
\centering
\caption{CVSS v3 metrics}
\begin{tabular}{|l|l|} 
\hline
\multicolumn{2}{|c|}{CVSS V3}  \\ \hline
Metric  & Metric Values      \\\hline\hline
Attack Vector (AV) & Network (N), Adjacent (A),
\\ & Local (L), Physical (P) \\\hline
Attack Complexity (AC)   & Low (L), High (H)  \\ \hline
Privileges Required (PR) & None (N), Low (L), High (H)   \\ \hline 
User Interaction (UI)    & None (N), Required (R)  \\ \hline
Scope (S)                & Unchanged (U), Changed (C)  \\ \hline 
Confidentiality (C)      & High (H), Low (L), None (N)\\ \hline
Integrity (I)            & High (H), Low (L), None (N) \\ \hline
Availability (A)         & High (H), Low (L), None (N) \\ \hline
\end{tabular}
\label{table:CVSS-v3}
\end{table}
\end{comment}
The NVD is the `U.S. government repository of standards based vulnerability management data' \cite{NVD}. It provides CVSS vectors and base scores for all vulnerabilities listed in the Common Vulnerabilities and Exposures (CVE) \cite{baker1999CVE} \cite{CVE} catalog of publicly disclosed software flaws. We use NVD to evaluate both CVSS v2 and v3 vectors and scores. The v2 data covers all CVE vulnerabilities published between 2005 and 2019. The v3 data ranges from 2015 to 2019 (only limited v3 data is available prior to 2015). These coverage dates result in the inclusion in our study of 118\,173 v2 vectors and scores and 55\,441 v3 vectors and scores. 

%********************************************************************

\section{Data Analysis}
\label{sec:Data-Anal} 

We analyze the NVD CVSS data in order to better understand the software vulnerability landscape. We investigate both the current nature of the threat posed by the existence and public disclosure of these vulnerabilities as well as how this threat has changed over time. To achieve this, we conduct the three studies described previously where we analyze the following: score distributions, metric value distributions, and relative rankings of the most frequent metric values.
%, and the most prevalent patterns of co-occurrence of the metric values.

  %---------------------------------------
        
\subsection{Score Distributions}
\label{sec:ScoreDistributions}

\begin{figure}[htp]
    \centering
    \includegraphics[width=0.45\textwidth]{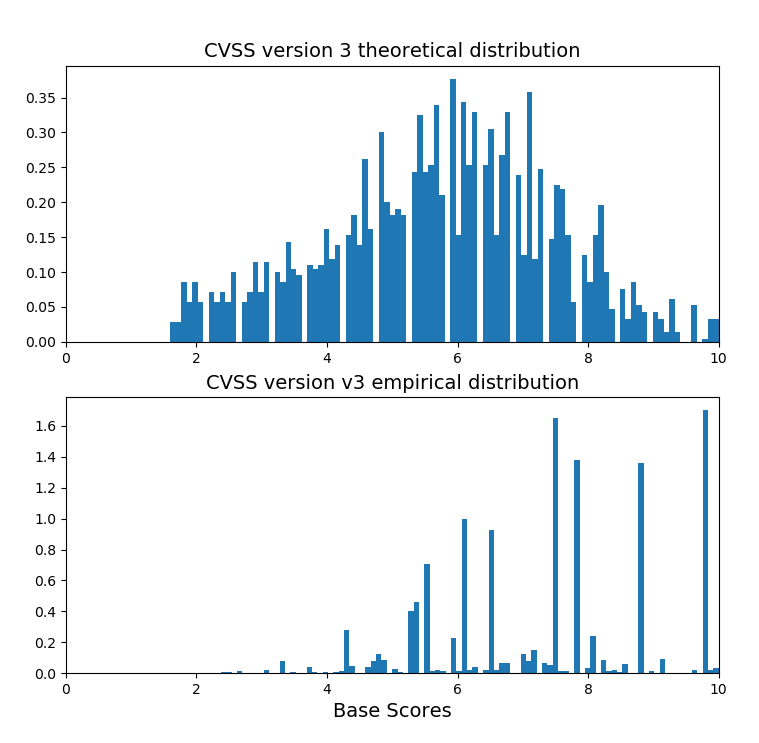}
    \caption{Theoretical vs Empirical Score Distributions for CVSS version 3}
    \label{fig:Theo_vs_Emp}
\end{figure}

The top graph of Figure \ref{fig:Theo_vs_Emp} shows the theoretical distribution of the v3 scores (V2 scores are similar and not shown in the paper due to space limitation. They can be found in the appendix of \cite{OurArxivVersion}). These plots show what is expected if all CVSS vectors (i.e., vulnerability types) are equally likely to occur. Note how the theoretical distribution was designed, by the FIRST CVSS committee, to spread CVSS scores throughout the range with a somewhat normal distribution with the most probable scores occurring in the middle of the distribution (a little biased to the right). That said, it is interesting in that for both v2 and v3 some scores are not possible even though they lie within the valid range of score values.
%(this is more prominent with v2 and only occurs with very low values in v3).

However, the empirical distribution is shown in the bottom  of Figure \ref{fig:Theo_vs_Emp} for v3. The empirical data indicates a predominance of certain vectors (groupings of vulnerability characteristics) in the real world. Thus, only a few vulnerability feature sets describe the majority of publicly disclosed vulnerabilities. This leads to the frequent use of just a very small number of scores. A similar observation was made in a previous study of the v2 scoring system \cite{mell2007improving}. 

The results observed with v3, which uses data from 2015 to 2019 (since v3 vectors are not generally available prior to 2015) are similar to those with v2, which uses data from 2005 to 2019. Hence, the long-term obtained with CVSS v2 data is confirmed by the shorter-term data of CVSS v3.

%The v2 distribution is comprised of NVD v2 vectors from 2005 to 2019. The v3 distribution is comprised of NVD v3 vectors from 2015 to 2019 (since v3 vectors are not generally available prior to 2015). We checked to see if the distributions changed over time and did not find any significant differences or trends; for this reason we just provide these two distributions using all available data. 

%One could argue that this was slightly remediated in v3 as v2 has 6 peaks above .2 while v3 has 8. However, the differences are relatively minor and both CVSS v2 and v3 .

  %---------------------------------------

\subsection{Metric Value Distributions}
\label{sec:ValueDistributions}

\begin{figure*}[htp]
    \centering
    \includegraphics[width=0.9\textwidth]{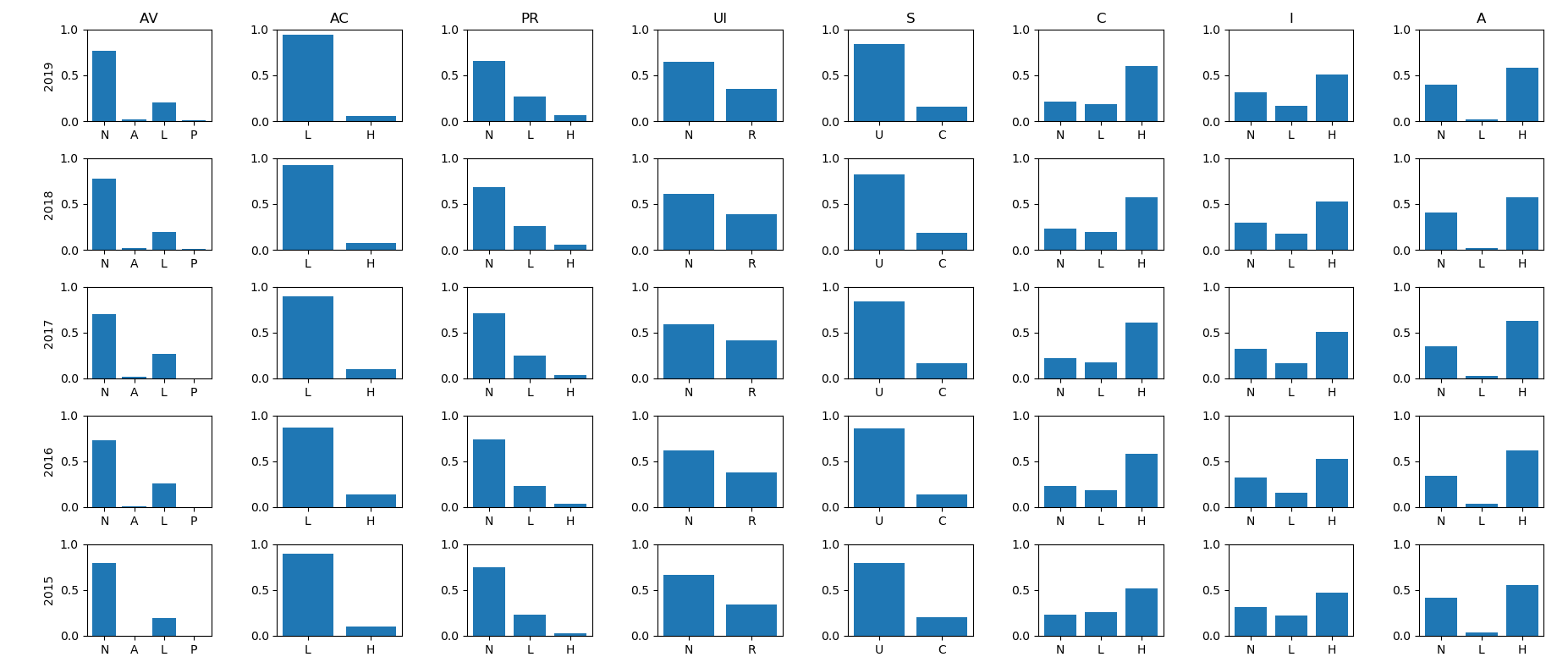}
    \caption{CVSS v3 metrics' values distributions over the years}
    \label{fig:CVSS3-Freqs}
\end{figure*}

To investigate more carefully (in order to identify) possible differences per year and trends over time, we focus on the distributions of each set of metric values per year over the time period of study. 
Figure \ref{fig:CVSS3-Freqs} provides the histograms for v3 from 2015 to 2019.
We have also plotted the histograms for v2 \cite{OurArxivVersion}, which cover from 2005 t0 2019. The inclusion of v2 in the study allows for a comparison over 15 years as opposed to being limited to just 5 years with v3, due to its more recent development. 

%\begin{figure*}[htp]
%    \centering
%    \includegraphics[width=0.9\textwidth]{images/CVSS_V2_Freq_Years_15_19.png}\\
%    \includegraphics[width=0.9\textwidth]{images/CVSS_V2_Freq_Years_10_15.png}\\
%    \includegraphics[width=0.9\textwidth]{images/CVSS_V2_Freq_Years_05_09.png}
%    \caption{CVSS v2 metrics' values distributions 2005-2019}
%    \label{fig:CVSS2-5_19_Freqs}
%\end{figure*}

The histograms for individual metric values for v3 appear almost the same year to year for the 5 years of study. This is the same in v2 over the longer time period of 15 years with some small exceptions: in 2014 the attack vector (AV) value of adjacent had some significance\footnote{According to the NVD team (in an email received March 10, 2020), this was a one time anomaly due to more than 800 CVEs all being announced simultaneously by an organization doing analyses on phone apps.}, the attack complexity (AC) value of medium increased some from 2007 onwards but then was steady, the authentication (Au) value of single increased slightly over the years, and the confidentiality (C), integrity (I), and availability (A) metric proportions between None, Partial, and Complete varied slightly from year to year while generally maintaining themselves about the same.

Overall though, the software vulnerability landscape for publicly disclosed vulnerabilities has been almost static during the period of study. This said, doing comparisons between the v2 and v3 histograms we see some differences, but this is due to differences in the approaches of the two versions of CVSS. These differences are primarily seen in regards to the metrics C, I, and A, which we will discuss shortly.

Consider the AV metric which reflects the context by which the vulnerability can possibly be exploited: Network (N), Adjacent (A), Local (L), or Physical (P). Both data sets show a high peak at N, a low peak at L and almost nothing at A and P. This indicates that the overwhelming majority of publicly disclosed software vulnerabilities are exploitable over the network (i.e., remotely), and it has been that way consistently through the period of study. 

The AC metric describes the conditions beyond the attacker’s control that must exist in order to exploit the vulnerability. When it is low (AC:L), the attacker can expect repeatable easy successes, while when it is high (AC:H) the attack is less likely to be successful. The data shows that the AC metric is largely dominated by the values of AC:L for v3 and AC:L and AC medium (AC:M) for v2.  This indicates that the set of publicly disclosed vulnerabilities have been predominantly easy to exploit.

This ``easiness'' to exploit vulnerabilities is confirmed by the other metrics for each CVSS version. For v3, the Privileges Required (PR) metric describes the level of privileges an attacker must possess before successfully exploiting a vulnerability. The user interaction (UI) metric captures the requirements for a human user (other than the attacker) to participate in the successful compromising of the vulnerable components. The data shows that in most of the cases, no privilege is required and very little user interaction is needed for a successful attack.

Similarly, with v2, the Au metric measures the number of times an attacker must legitimately authenticate to a target in order to be in a position to exploit a vulnerability. The data shows that almost always, there is no authentication required prior to exploiting a vulnerability. Sometimes a single authentication is required, but almost never is there a vulnerability that requires multiple authentications in order to be successfully exploited. 

CVSS v3 introduced a new scope (S) metric, which captures the spill-over effect: how much a vulnerability in one vulnerable component impacts resources in components outside of its security scope. 
When the scope is unchanged (S:U), there is no spill-over, while when the scope is changed (S:C) the vulnerability will very likely affect other components. The data shows that the scope metric has predominantly been S:U.

The last three metrics C, I, and A are common to both CVSS versions. They capture the extent to which a successful exploitation of a vulnerability will affect these three principles of security on the effected component.
With respect to these metrics, the v3 data shows that the impact on C, I, and A has predominantly been high (C:H, I:H, and A:H) with very similar distributions for all the years. The v2 data also shows a similar stationary behavior in the distributions. However, the difference in the fraction of high for v3 and complete for v2 is notable.
%, specifically for the A metric. 
One might expect the high values in CVSS v3 to be equivalent to the complete values for v2. However, this is not the case as they are defined differently. 

%in an email exchange with the NVD team in October 2019, they said that ``the CVSS scoring systems are fundamentally different regarding qualifications for CIA Partial/Complete and Low/High. This is a common misconception due to the nuances of the scoring systems. In addition to this, the NVD takes the approach of representing the worst-case scenario when information is lacking. This typically results in default values of HIGH being attributed to a CVE unless data is available that identifies a limitation to the impact or meets qualifying text for the specification.''

  %---------------------------------------

\subsection{Relative Rankings of the Most Frequent Metric Values}
 \label{sec:RelativeRankings} 

\begin{figure*}[htp]
    \centering
    \includegraphics[width=0.9\textwidth]{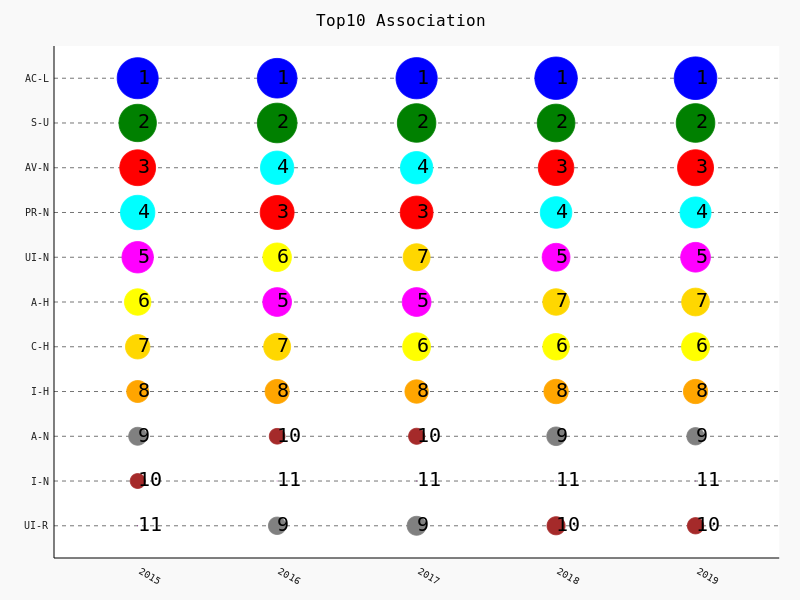}
    \caption{CVSS v3 top 10 rankings}
    \label{fig:CVSS3-Top10}
\end{figure*}

We now focus on the most prominent individual values of the metrics, evaluating the rankings of the top 10 metric values observed each year and providing a comparison between the years. Figure \ref{fig:CVSS3-Top10} shows the rankings for v3 (the same plots for v2 can be found here \cite{OurArxivVersion}). The y-axes show the top 10 most prevalent metric values, ordered from the least frequent to most frequent. Thus, the set of metric values included in the y-axis is significant (only the top ten are shown). The x-axes show the years. Each (\textit{x},\textit{y}) point indicates that on year \textit{x} the metric value at \textit{y} has a rank indicated by the number in the circle. The size of the circle is proportional to the number of times that metric value appeared in a score in that year. For example in Figure \ref{fig:CVSS3-Top10}, in 2017 the metric value AV-N was the fourth most frequent metric value within the set of all v3 vectors. However, in 2018 and 2019 this metric value became the third most frequent. Notice that in general, a value might appear in the top 10 of one year while not appearing in another year. Whenever that happens, we rank that particular value 11 for all the years in which it did not appear.

For v3 (see Figure \ref{fig:CVSS3-Top10}), we observed that the same top 10 values appeared from 2016 to 2019. Furthermore, only one of those values is missing in the 2015 top 10. In addition, these values were ranked almost the same over the years. The top 2 are constant and in the same order over the time period 2015 to 2019. The top 4 and the bottom 4 (including the 11th appended value) are also constant with minor changes in the order they appear over the years. The v2 data shows similar results (see \cite{OurArxivVersion}). This is another illustration of the stationary threat landscape observed earlier. It also corroborates the observations in Figure \ref{fig:Theo_vs_Emp}, that the landscape has been dominated by just a few vulnerability types.

%\begin{figure*}[htp]
%    \centering
%    \includegraphics[width=0.9\textwidth]{images/CVSS_v2_Top10_1.png}
%    \caption{CVSS v2 top 10 rankings}
%    \label{fig:CVSS2-Top10}
%\end{figure*}

In conclusion, our data indicates that the vulnerability threat landscape has been dominated by a few vulnerability types and has not evolved over the years. The overwhelming majority of software vulnerabilities are exploitable over the network (i.e., remotely). The complexity to successfully exploit these vulnerabilities is dominantly low and very little authentication to the target victim is necessary for a successful attack. Moreover, most of the flaws require very limited interaction with users. The damage of these vulnerabilities has however mostly been confined within the scope of the compromised systems.

\section{Related Work}
\label{sec:relatedWork} 

There are many efforts to understanding the software vulnerability landscape. These efforts include reports by security solutions vendors \cite{SymnatecReport2019,MCAfeeReport2019}, white papers from non-profits such as MITRE \cite{MitreTop25} and SANS \cite{SANSCyberThreat}, as well as academic papers\footnote{Any mention of commercial products or entities is for information only; it does not imply recommendation or endorsement.}. 
%There are few studies that use existing standards, datasets, and databases, such as the NVD, CWE, CVE, ExploitDB, and CVSS. 
For CVSS, most studies focused on the aggregation equation that produces the CVSS numerical scores representing the severity of the vulnerability. Surprisingly, we found no studies on v3 despite its preponderance in commercial security software.
%In this paper, we use the raw CVSS vectors (not the numerical scores) to perform historical and statistical analysis of the threat landscape. We consider both v2 and v3. We did not attempt to make prediction on the future of threats. However, our study can serve as a basis for understanding future threats.

Reference \cite{mell2007improving} is among the first statistical studies of the CVSS scoring system. It evaluates v1 and proposed improvements that contributed to the release of v2. Our study considers both v2 and v3 (but doesn't try to improve on either). Relative to the statistical evaluation, we consider our paper as a continuation and update of the work in \cite{mell2007improving}. However, our work uses data from a much longer time period. It also goes one step further by analyzing association rules of vulnerability metrics. It is worth noting that there are similarities between the results of the two studies. For instance, both papers show the predominance of certain types of vulnerabilities. Our temporal analysis (which was not performed in \cite{mell2007improving}) shows that this predominance is maintained over the years. 

Reference \cite{scarfone2009analysis} considers CVSS v1 and v2 and analyzes how effectively v2 addresses the deficiencies found in v1. It also identifies new deficiencies. In contrast, our motivation was to understand the threat landscape.

Reference \cite{holm2012empirical} uses empirical data from an international cyber defense exercise to study how 18 security estimation metrics based on CVSS correlate with the actual \emph{time-to-compromised}  (TTC) of 34 successful attacks. This study uses TTC  as a dependent variable to analyze how well different security estimation models involving CVSS are able to approximate the actual security of network systems. The results suggest that security modeling with CVSS data alone does not accurately portray the time-to-compromise of a system. This result questions the applicability of the CVSS numerical scoring equation. Our study focused on the raw CVSS vectors, which represent the actual experts’ opinions about the vulnerabilities.

%Another empirical study of the CVSS was performed by Zhang \emph{et. al.} \cite{zhang2011empirical}. Similar to our paper, 
Reference \cite{zhang2011empirical} uses NVD data to study trends and patterns in software vulnerabilities in order to predict the time to next vulnerability for a given software application. Data mining techniques were used as prediction tools. The vulnerability features used to aid the prediction are the published time of each vulnerability and its version. We believe that these features are not sufficiently informative. Instead, we directly use the eight metrics from the CVSS base scores which constitute the best available information covering large multi-year sets of vulnerabilities.

Reference \cite{Feutrill2018} also carried out a predictive study based on the NVD/CVSS and ExploitDB \cite{ExploitDB} data. Using the CVSS data, it attempts to answer two questions: \emph{ (1) Can we predict the time until a proof of concept exploit is developed based on the CVSS metrics? and (2) Are CVSS metrics populated in time to be used meaningfully for exploit delay prediction of CVEs?} The former is answered in the positive, while the latter is answered in the negative. While using the same datasets, our objective differs from that in \cite{Feutrill2018}. We did not attempt to predict the threat landscape; we provide a thorough historical and statistical study of vulnerabilities for the last fifteen years.

The work in \cite{johnson2016can} is another assessment of CVSS. It evaluates the trustworthiness of CVSS by considering data found in five vulnerability databases: NVD, X-Force,
OSVDB (Open Source Vulnerability Database), CERT-VN (Computer Emergency Response Team, Vulnerability Notes Database) , and Cisco IntelliShield Alerts. It then uses a Bayesian model to study consistencies 
%(e.g., public vulnerability information on which to base the scoring) 
and differences. 
%(e.g., scorers’ background knowledge)
It concluded that CVSS is trustworthy and robust in the sense that most of the databases generally agree. This suggests that our focus on the NVD to study the threat landscape is justified: studies using data from the other databases will likely lead to the same conclusions.

All of the studies cited above are focused on v1 and v2. In our literature survey, we did not find a single study that uses the updated and significantly modified v3 to understand the software vulnerability landscape. We believe that the present paper is the first of this kind in doing so. Furthermore, our study is the first to use association rule mining and co-occurrence of vulnerability metrics' values in an attempt to characterize the software threat landscape. 

%********************************************************************

\section{Conclusion}
\label{sec:conclusion}

%In this paper, we undertook the collection and analysis of CVSS v2 and v3 data sets. Our analyses focused on four main areas of interest. First, we analyzed the empirical and theoretical distributions of v2 and v3 scores. Second, we analyzed the distribution of CVSS metric values and how vulnerability characteristics change over time. Third, we studied the most frequent metric values and analyzed their rankings over a period of 15 years. Fourth, we used association rule mining techniques to derive the most prevalent patterns of co-occurrence of the metrics.

Our data indicates that the vulnerability threat landscape for publicly disclosed vulnerabilities has been dominated by a few vulnerability types and has not significantly changed from 2005 to 2019. However, the underlying software flaw types that enable these vulnerabilities change dramatically from year to year (for example, see \cite{NVDCWE}). This means that many flaw types result in vulnerabilities with the same properties. This is bad news because it means, as a security community, it will be difficult to eliminate certain vulnerability types because they result from a plethora of underlying software flaw types.

Another concern is that the overwhelming majority of software vulnerabilities are exploitable over the network. When developing software, efforts should be made to reduce unnecessary connections, protect necessary ones, and require more authentication where possible to reduce attack surface area. Another significant issue is that most of the vulnerabilities require no sophistication to be exploited (but again this is hard to improve upon due to the many software flaw types that allow this).

These two factors together combined with the finding that most vulnerabilities require very limited interaction with users facilitates the widespread hacking occurring today. Often in security literature the human is cited as the weakest link. While certainly humans can be exploited, within the set of CVE type vulnerabilities exploitation of humans plays a very minor role; training humans will have little impact in this area.

%The complexity to successfully exploit these vulnerabilities is dominantly low and very little authentication to the target victim is necessary for a successful attack. Moreover, most of the flaws require very limited interaction with users. The damage of these vulnerabilities has however mostly been confined to the scope of the compromised systems with very moderate spill-over. Lastly, we identified a variety of associations such that the appearance of certain vulnerability attributes implies the existence of another.

Overall, this study documents the security community's inability to eliminate any types of vulnerabilities through addressing the related software flaw types. In 15 years, the vulnerability landscape hasn't changed; through the lens of the metrics in this paper we aren't making progress. Perhaps we as community need to ``stop and think'' about the ways we are developing software and/or the methods we use to identify vulnerabilities. The security community needs new approaches. We don't want to write this same paper 15 years from now showing that, once again, nothing has changed.

Overall, this study shows that either we (the community) are incapable of correcting the most common software flaws, or we are focusing on the wrong flaws. In either case, it seems to us that there is a need to ``stop and think'' about the ways we are developing software and/or the methods we use to identify vulnerabilities.

%********************************************************************
\begin{comment}
\section*{Acknowledgement}
This work was partially accomplished under NIST (National Institute for Standards and Technology) Cooperative
Agreement No.70NANB19H063 with Prometheus Computing, LLC. The authors would like to thank the NVD staff for their review and consideration of this work.
\end{comment}
%********************************************************************

\bibliographystyle{IEEEtran}
\bibliography{IEEEabrv,MCbody}

\begin{appendix}

The figures in this appendix show the CVSS v2 metrics' value distributions for the 15 years from 2005 to 2019.

\begin{figure}[htp]
    \centering
    \includegraphics[width=0.45\textwidth]{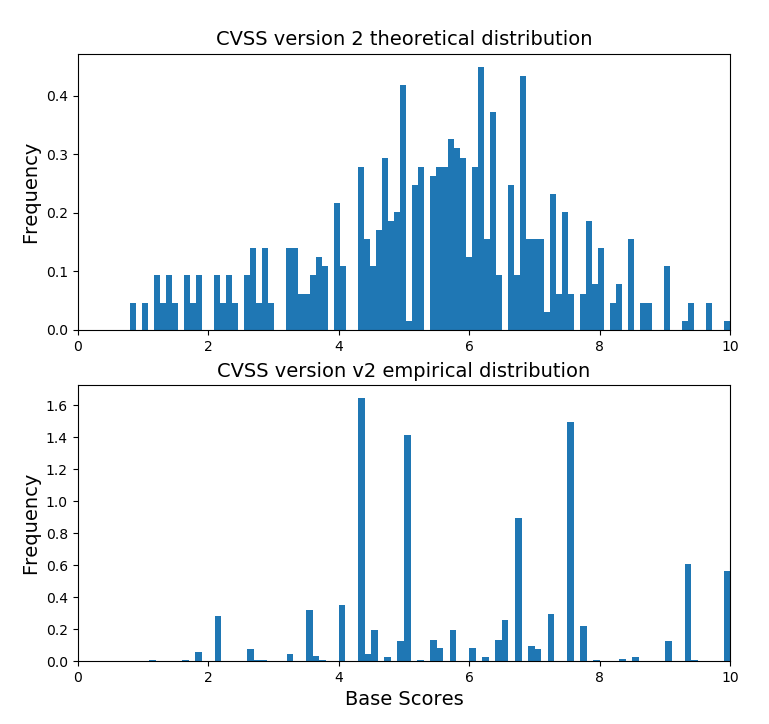}
    \caption{Theoretical vs Empirical Score Distributions for CVSS version 2}
    \label{fig:Theo_vs_Emp}
\end{figure}

\begin{figure*}[htp]
    \centering
    \includegraphics[width=0.9\textwidth]{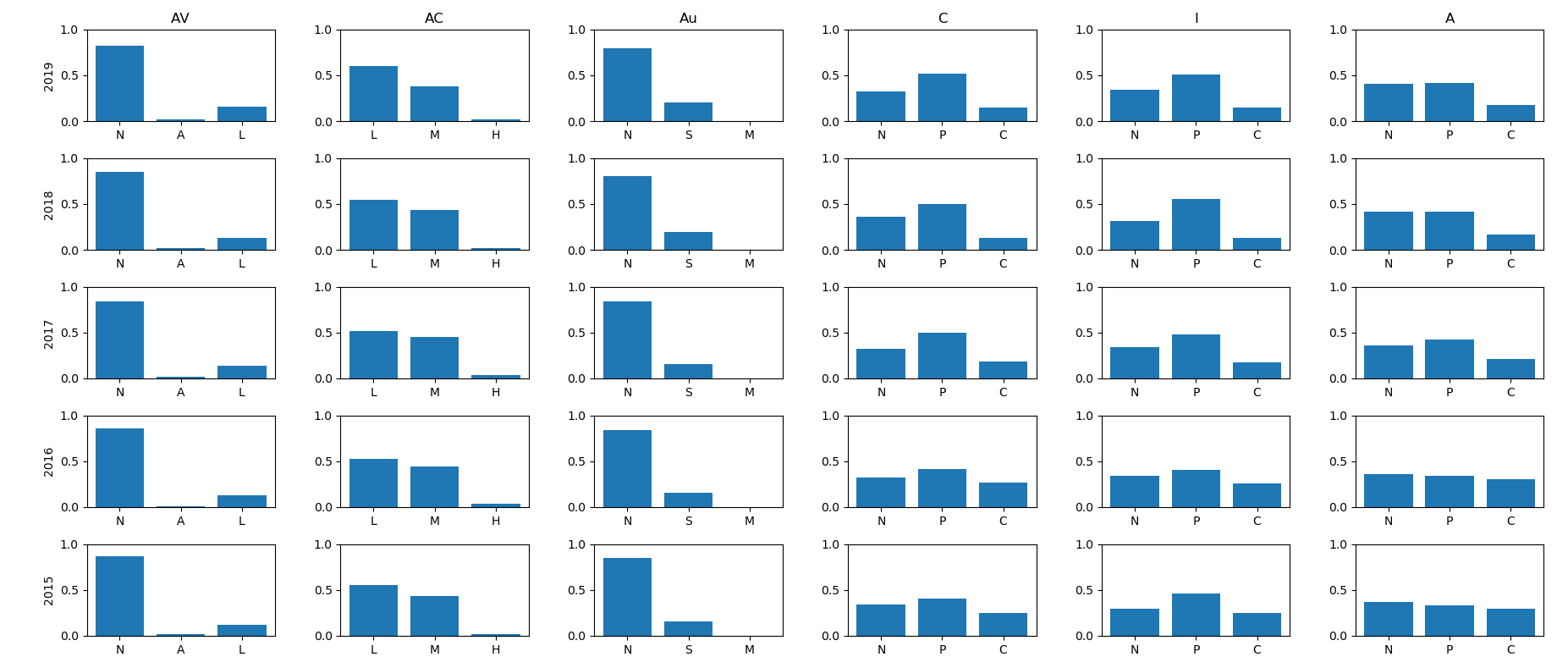}\\
    \includegraphics[width=0.9\textwidth]{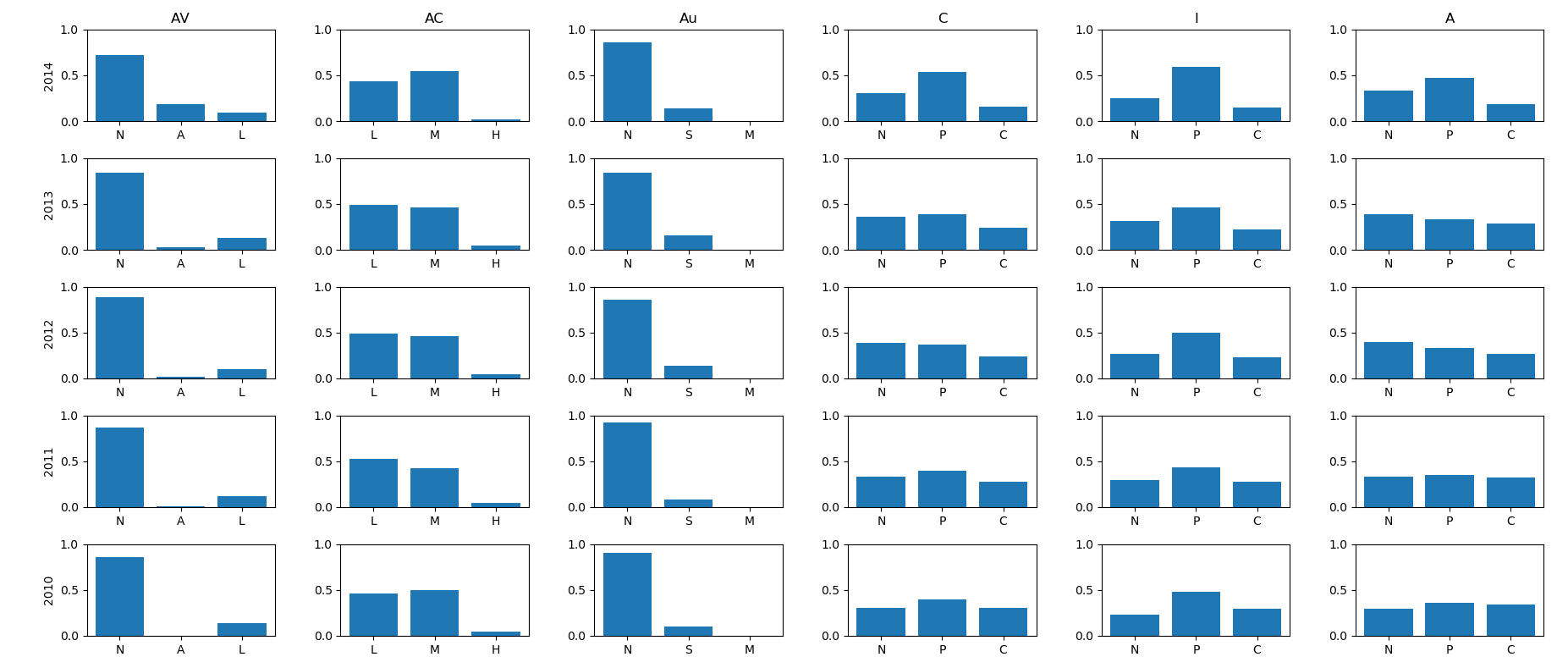}\\
    \includegraphics[width=0.9\textwidth]{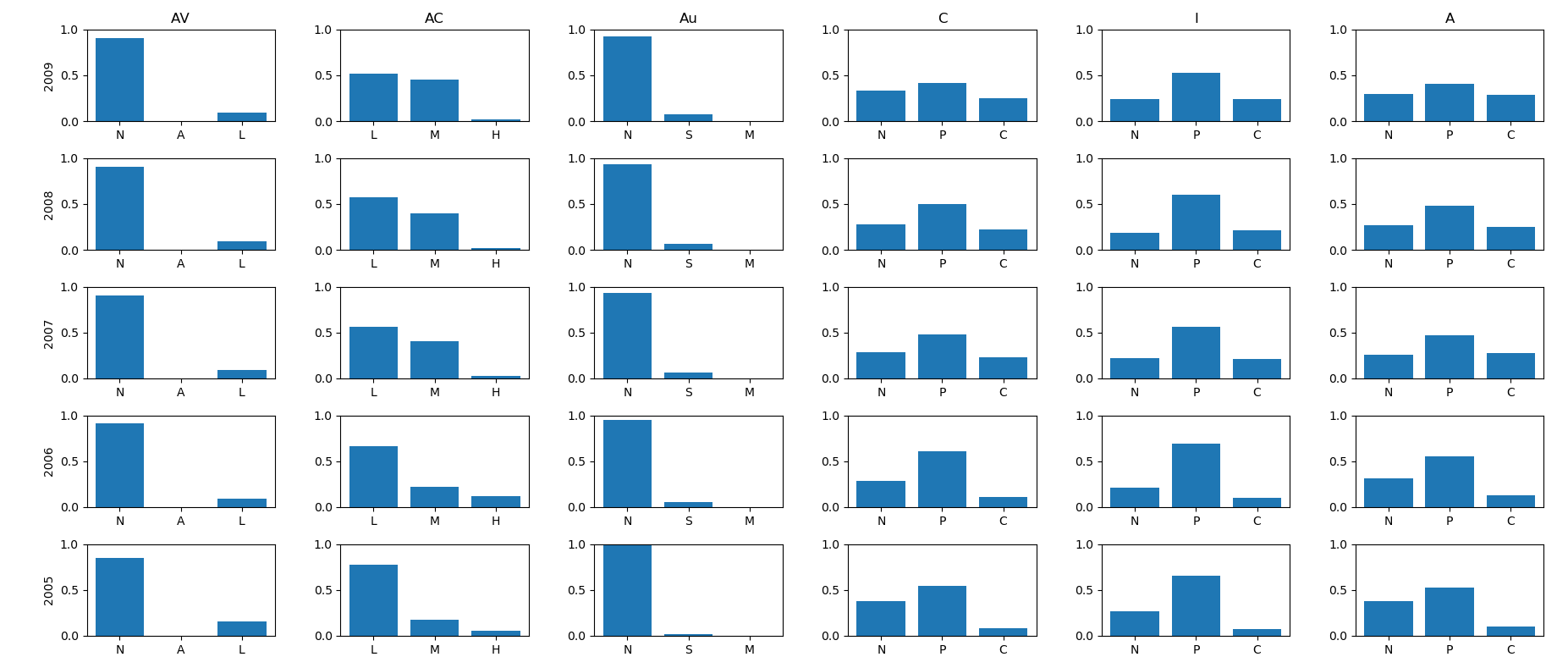}
    \caption{CVSS v2 metrics' values distributions 2005-2019}
    \label{fig:CVSS2-5_19_Freqs}
\end{figure*}

\begin{figure*}[htp]
    \centering
    \includegraphics[width=0.9\textwidth]{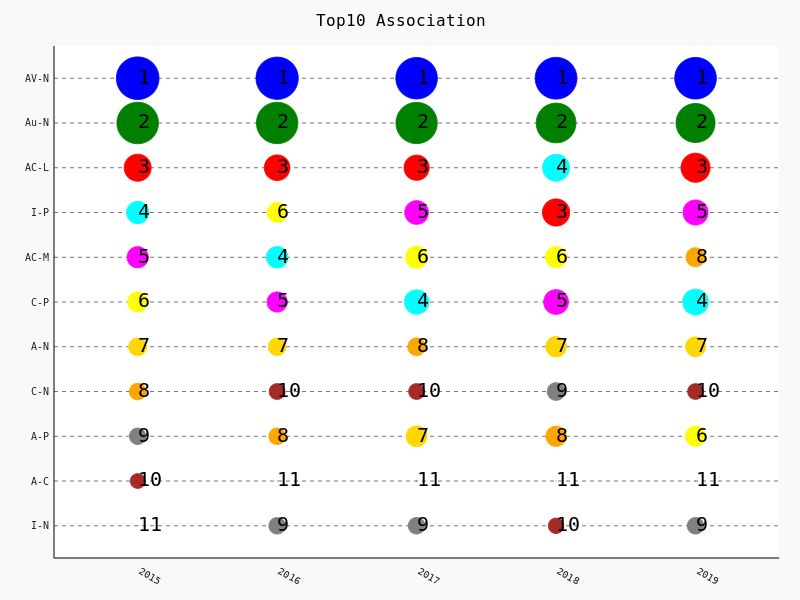}
    \caption{CVSS v2 top 10 rankings}
    \label{fig:CVSS2-Top10}
\end{figure*}

\end{appendix}

\end{document}